\documentstyle[12pt]{article}
\begin{document}

\begin{center} {\bf DIFFUSION ON A REARRANGING LATTICE}  \\
Aninda Jiban Bhattacharyya$\footnote[1]{email: aninda@juphys.ernet.in}$ 
and S.  Tarafdar
 \\Condensed Matter Physics 
Research Center,\\ Department Of Physics, Jadavpur University,\\
Calcutta-700032, INDIA. \end{center}
\begin{abstract} In this paper we present a computer simulation of a  
random walk
(RW) for diffusion on a rearranging lattice. The lattice consists of two  
types of 
sites -- one good conducting (type 1) and the other poor  conducting  
(type 2),   distributed at random. The two types of sites  
are assigned different waiting times ($\tau_{1}$ for type 1 and 
$\tau_{2}$ for  type 2) . We assume that at intervals of time $\tau_{r}$, 
the site distribution changes. The effect of this rearrangement on  the 
diffusion 
coefficient is studied with varying $\tau_{r}$. We study this effect  for 
different 
ratios of dwell times of the two types of sites (R) and also for different
fractions (X)  of the less conducting sites. An empirical relation for  
D($\tau_{1}$
,$\tau_{2}$,$\tau_{r}$,X) is suggested. We have employed the well model 
and considered diffusion controlled by sites, rather than bonds. So our  
approach is different 
from the dynamic bond percolation model, which studies these aspects.  
Our results show that the diffusion coefficient D may  change 
by a factor of upto 3 (approximately) for rapid rearrangement, and there is a 
considerable effect of varying X and R on the range of variation of D,
where X is the fraction of low conducting sites, and R is the ratio of the
dwell times for the two types of sites. Further
 for $\tau_{r}$ > 250$\tau$ ($\tau$ is the time unit for the random walk) 
the effect of rearrangement
becomes negligible.  The results may be useful for studying diffusion and 
conduction  of ion conducting polymers.
\end{abstract} {\bf PACS NO : 66.30.Dn. ; 61.43.Bn.
}
\newpage
\section  {\bf Introduction : } 

A convenient way to study diffusion and conduction is through the  random 
walk
 formalism. Disordered media, which comprise many systems of practical 
interest have 
 been analytically treated by this method [1-3]. However, an additional 
complication
m ay be present in the system in the form of dynamic disorder. In this 
case the system
undergoes some reorganisation with a characteristic timescale. This 
picture applies to
polymers and also to glasses above their  glass transition temperature.
Exact analytical results for one version of this situation have  been 
obtained
 in a series of works [4-6], where the dynamical bond percolation model 
(DBPM) has been proposed and treated in detail.

In the present communication we are interested in a different approach  
to the
same problem. We consider a "well" model instead of the "barrier" model.
In other words we are concerned with {\it sites} rather than bonds. 
 We study diffusion in this system in two-dimensions through a computer 
simulation of the random walk including the rearrangement effect.

 No exact results are available for this model with finite renewal time 
$\tau_{r}$
though the limits for $\tau_{r} = 0$ and $ \tau_{r} = \infty$ can be exactly 
 calculated. We have varied $\tau_{r}$ from $\tau $ to $ 250.\tau$, where 
$\tau$ is
the time unit for the random walk. Our simulation results agree with the
 limiting values and we show that variation in D with $\tau_{r}$ within 
the limits
can be represented by a simple interpolation formula. We also find that 
 the limit for large $\tau_{r}$ is approached very closely for $\tau_{r} 
= 250\tau$.

We have
considered random arrangements of two types of sites with different
conductivities represented by different dwell times. 
This work has been developed with the quasi 2-dimensional polymer
thin films in mind; here the highly conducting sites represent the
amorphous phase and the low conducting sites the crystalline phase.
However the treatment is general and may be applied to any other situation
 where there are two different species with different conductivities and 
rearrangement.             
 It is our aim to identify clearly the regime where this effect is most
important and to assess how much it can affect the diffusion coefficient.
We show that the diffusion coefficient may increase by a factor of about
3  for very rapid rearrangement, i.e. if the characteristic rearrangement
time $\tau_{r}$ is of the same order as the hopping time $\tau$. But for
$\tau_{r} > \tau_{s} \approx 250\tau$ there is no significant change in the
diffusion coefficient (D) i.e. the diffusing particle sees the lattice as
a quenched disordered system. 

We have studied D versus $\tau_{r}$ curves varying two parameters -- (a)
the ratio of dwell times on the two different types of sites 
 \begin{equation} R = \frac {\tau_{2}}{\tau_{1}} = \frac{p_{1}}{p_{2}} 
\end{equation} $\tau_{1}$, $\tau_{2}$, p$_{1}$ and p$_{2}$ 
 are the dwell times and jump probabilities of type 1 and 2 sites 
respectively, and (b) the fraction of low conducting sites X. We 
also measure the range of variation in D for different X. This
may be measured by  $\Delta$ D = D$_{max}$ - D$_{min}$  or 
 D$_{r}$ = D$_{max}$/D$_{min}$. $\Delta$D has a maximum at X=0.2, and 
D$_{r}$ is  maximum at X=0.5.

In the next section our random walk algorithm is described and in section
 3 the results are given. In the last section we discuss the implication 
of our
results and compare them with earlier work on related problems. \section
 {\bf The random walk algorithm :} 

Our random walk algorithm is an extension of the algorithm used earlier by 
Bhattacharyya et al [7], the new feature is that here $\tau_{r}$ is explicitly
specified and varied. As mentioned in the earlier section, the random walk
(RW) is performed on a 2-dimensional square lattice containing two types
 of sites. The distribution of the two different types in the lattice is
random. The lattice sites actually represent small regions of the system
which belong to a single phase only, and the lattice spacing ($\xi$)
represents the distance between such sites. A site belonging to the
i$^{th}$ phase is assigned a jump probability p$_{i}$ for jumping to a
nearest neighbour site, at each time step. This implies an average waiting
time $\tau_{i}$ at the i$^{th}$  site. A longer waiting time corresponds
to a lower conductivity  of the phase and hence of the site. 

A distribution of energetically different sites on a
lattice is usually represented in a simulation model by either of the
following pictures : \newline 1. Well model \newline 2. Barrier model 

In the well model each site is treated as a potential well. The well depth
w$_{i}$ is a characteristic of the site type i and it determines how long
the random walker will be trapped there. 

In the barrier model a barrier b$_{ij}$ is assumed to exist between sites
i and j. The probability for the particle to hop from i to j is determined
by the nature of both the sites i and j. 

In the present work we have employed the considerably simpler well model,
where the probability of jumping from a site is determined by the phase of
 that site only. The walker i.e. the diffusing particle  is allowed to go 
to any of
the four nearest neighbours with equal probability whether they belong to
the highly or poorly conducting phase. We find normal diffusion as expected
with the mean square distance travelled proportional to the time. 

The steps of the RW are as follows.
\begin{enumerate} 
\item The walker starts on an initial lattice point (x$_{o}$,y$_{o}$) on a
 two dimensional lattice. A random number is chosen to determine type of 
the site,
and another to determine whether it jumps and its final position after the jump.

\item We assume that the host lattice retains the memory of a
certain  distribution of sites for $\tau_{r}$ time steps, after that there
is a rearrangement. This is implemented as follows. 

During the time interval $\tau_{r}$ the walker stores the coordinates and
the character (whether type 1 or type 2) of the sites visited. So if the
 walker visits the same site more than once within the interval $\tau_{r}$,
it finds there the same phase as was present earlier. \item  Step (1) is
repeated  during the time interval $\tau_{r}$, with the current 
site coordinates instead of (x$_{o}$,y$_{o}$). \item At t = ($\tau_{r} +
1$), the system forgets the previously stored $\tau_{r}$ sites and their
character and starts a fresh list for the next interval. \item Steps (1-3)
are repeated again for the next time interval $\tau_{r}$. 
\end{enumerate}
The above procedure is repeated K times, where \begin{equation}
 K = \frac{N_{total}}{\tau_{r}} \end{equation} N$_{total}$ --- total 
number of time
steps for a particular walk. \newline Due to the stochastic nature of the
process, one has to average over a large number of such walks to get a
meaningful value of <r$^{2}$>. In this work the walker executes a random
walk of (15000-75000) steps and r$^{2}$ is averaged over (20000-100000)
walks. This gives sufficiently good convergence upto 3 significant figures
for the diffusion coefficient. We have calculated D for $\tau_{r}$ varying
from (1-250). For still higher $\tau_{r}$ there is negligible change in D.

This random walk algorithm allows the walker to move on an effectively
infinite sample. This is possible because here we do not take a quenched
system with sites assigned specifically to a definite phase. So the
problem of finite size effects is avoided. There is however
a restriction to the walk size due to limited computer time. 
\section {\bf Results  :} \subsection {\bf Variation in D with
$\tau_{r}$  for constant R and X : }

Using the above RW algorithm, the diffusion coefficient was obtained as a
 function of  $\tau_{r}$, X and R. We have kept $\tau_{r}$ constant at 
$\tau_{2}$ =10,
 and $\tau_{1}$ has been varied from 10 to 1.01. We find that D does not 
change 
 on further decrease of $\tau_{1}$. The minimum value $\tau_{1}$ can have 
is 1 corresponding to p$_{1}$ = 1.
Figs. 1 and 2 are plots of D versus $\tau_{r}$ for different
values of R (9.9,8,6 and 1) at X=0.14 and X=0.80 respectively. 
Fig.3 is plot of D versus
1/$\tau_{r}$ for different R values at a constant X(=0.14).
 The maximum and the minimum values for D corresponding to $\tau_{r} 
\rightarrow 
 $ 0 and $\tau_{r} \rightarrow \infty$ are given by \begin{equation} 
D_{max} = \frac{1}{4\tau_{1}} \left[ (1 - X) + \frac{X}{R} \right ]
\end{equation} and 
 \begin{equation} D_{min} = \frac{1}{4\tau_{1}} \left [(1 - X) + XR 
\right ]^{-1} \end{equation} with lattice constant = 1. These values agree with the simulation results.
In figures 1 and 2 the simulation results for D are shown as discrete points,
from the variation in D we suggest an empirical relation for $D=D(\tau_{r},
\tau_{1},\tau_{2}, X$). Calculated values from this relation are shown as
continuous curves.

We now give the proposed formula for D. D may be written in terms of an
effective time $\tau_{eff}$ which is some sort of average over the
 characteristic times for the two types of sites. So $ D= 1/4\tau_{eff}$. 
Casting equations (3) and (4) in this form, we have
\begin{equation}
 \tau_{eff}(\tau_{r} \rightarrow \infty) = \tau_{max} = \tau_{1}(1 - X) + 
\tau_{2}X \end{equation}
and
\begin{equation}
 \frac{1}{\tau_{eff}}(\tau_{r} \rightarrow 0) =\frac{1}{\tau_{min}} = 
\frac{1-X}{\tau_{1}} +\frac{X}{\tau_{2}} \end{equation}
    i.e. in the first case the walker sees an average waiting time, 
whereas in
the second case it sees an average jump frequency. In other words in the
first case we have a Voigt average of waiting times and in the second case a
Reuss average [8].
For finite $\tau_{r}$ we propose the following relation 
\begin{equation}
    \frac{1}{\tau_{eff}(X,\tau_1,\tau_2,\tau_r)} = \frac{1}{\tau_{max}}. 
\frac
 {\alpha_{2}}{\alpha_{1} + \alpha_{2}} + \frac{1}{\tau_{min}}.\frac 
{\alpha_{1}}
 {\alpha_{1} + \alpha_2}
\end{equation}
where
\begin{equation}
\alpha_{1} = 1 - exp \left (-\frac{\tau_{1}}{X \tau_{r}} \right )
\end{equation}
\begin{equation}
\alpha_{2}=exp \left (-\frac{\tau_{2}}{(1-X)\tau_{r}} \right )
\end{equation}
This relation reduces to the correct limits for X=0 and X=1, as well as
$\tau_r\rightarrow 0$ and $\tau_r \rightarrow \infty$. It reproduces quite
 well the strong nonlinearity in D between the limiting values. The 
calculated
values are slightly higher than the simulation results, as shown in figures
1 and 2.  The agreement between the empirical formula and simulation
results is better for low X. For higher X the calculated D does not fall
as sharply with $\tau_{r}$, as the simulation results.

Salient features of our results are as follows.
The change in D with $\tau_{r}$ is
significant for $\tau_{r} < \tau_{s} \approx 250\tau $. The hopping time
$\tau$ is the smallest time scale for our system. For $\tau_{r} >
    \tau_{s}$ the system is effectively quenched. This is seen more 
clearly in
 fig.3. We find that the limiting value of D for $\tau_{r} \rightarrow 
\infty$
 is hardly different from the last calculated data point corresponding to 
$\tau_{r} =
250\tau $. So we take $\tau_{s}$ = 250$\tau$. For small X and large R, i.e. 
 the residence probability of the type 1 phase being negligibly small, 
the average time elapsed between  jumps 
$\tau_{i}$ may be approximated as $\tau$. In this case 
we can  
 get an estimate of $\tau_{s}$ in real units. For a polymer system, 
$\tau_{i} \approx
 10^{-6}$ sec according to nuclear magnetic resonance (NMR) linewidth 
narrowing 
experiments [9,10]. so from $\tau_{s}=250\tau$, we find that $\tau_{s} \sim
10^{-4}$ sec.
\subsection  {\bf Range of variation of D  for different X : }

 It is obvious that for any value of R, $\Delta $D = 0 and D$_{r}$=1 for 
both X=1 and 
X = 0, i.e. if there is only one type of site on the lattice. We find that
 D$_{r}$ has a maximum for X = 0.50, whereas $\Delta$ D has a maximum  at 
X = 0.20.
 This is because D itself is larger at lower X. The X values of peaks 
obtained for D may not correspond to 
those obtained while studying conductivity. This is due to the fact that
the charge carrier concentration comes into play when conductivity is
calculated. 

 Again if R = 1, i.e. all sites are equivalent, $\Delta$D=0 and 
D$_{r}$=1. The 
effect of increasing R towards R $\rightarrow \infty$  is seen in figs 1 and 2.
\section {\bf Discussion :}

\subsection{Comparison with DBPM Model :}
   We discuss briefly the DBPM model which is concerned with the barrier 
model i.e. the bond picture of the same problem.

Starting with a one-dimensional model, a series of works have been published
[4-6] which develop the dynamic bond percolation (DBPM) model, including 
different features and extending it to higher dimensions. These works
develop an analytical approach to the problem of diffusion in a rearranging
lattice with bond renewal. Here one type of site is conducting and the other
 is completely insulating. The review by Nitzan and Ratner [4] gives a 
complete overview of the model. 

The most significant result of this work is the demonstration that, the 
diffusion
coefficient D($\tau_r$) with renewal is  identical to a frequency
 dependent diffusion coefficient D($\omega$) on a static lattice through 
 an analytical continuation
 rule.$$D(\omega,\tau_{r})=D_{o}(\omega-\frac{i}{\tau_{r}})$$ Their work 
is also compared with effective medium models [4].

 The present model for $R \rightarrow \infty$ may be compared with the DBPM.
 A basic difference is to be noted in the two cases. In DBPM or any
standard bond percolation model the insulating sites are blocked, that is
inaccessible to the walker. In the present model, however, the insulating 
sites are infinite traps, from which the walker cannot escape. The diffusion
behaviour of the two models is quite similar, in spite of this difference.
Let us consider the situation below the percolation threshold. In DBPM the
walker gets confined to a finite cluster after some characteristic time
which is a function of X. If $\tau_{r}$ is larger than this time, the mean
square distance travelled saturates to a constant value. On renewal , the
sites rearrange and the walker is released from the previous cluster it
occupied. Now $<r^2>$ starts to increase again. This continues in steps as 
shown in fig(5) of ref [5].

Let us now consider a similar situation in our model. Here the walker gets
trapped in one of the insulating sites after a certain time depending on X.
But after renewal the site may change to a conducting site and release the
 walker, so $<r^{2}>$ increases in steps just as in the DBPM. This shows 
that the
overall diffusion behaviour is similar, though the microscopic pictures are
quite different. Of course the percolation threshold in this case, is 
expected to be the site
percolation rather than the bond percolation threshold appropriate for the
DBPM.
We have not yet attempted simulation of this limiting situation.

 The qualitative appearence of the curves for 
 D vs $1/\tau_r$ in ref[5] is very similar to our figure 3, which 
illustrates the underlying similarity of the two approaches.

\subsection{Conclusion :}

The effect of rearrangement of the lattice due to liquid-like behaviour
at short length scales is considered to be very important for studying
conduction in polymers. We have assessed how important it can be and when
particularly it must be taken into account. 

This effect becomes unimportant after $\tau_{r} >
\tau_{s}$. Assuming $\tau \sim 7 \times 10^{-7}$ sec [9,10],
$\tau_{s}$ is of the order milliseconds. For larger $\tau_r$ it suffices to
take the quenched lattice limit.

Our system is an infinite
 lattice, so finite size effects which may distort the results considerably
 are absent. It is to be noted that our definition of $\tau_{r}$ refers  to
 the time for interchange of crystalline and amorphous sites. This is similar
 to the  original definition of renewal time ($\tau_{ren}$) by Druger et 
 al [5], but in
 other works different renewal times for  crystalline and amorphous regions
 have been considered [11]. Chang and Xu have considered rotation of 
 polymer chain sections in their work [12]. 

 We plan to incorporate our findings reported here, in an ongoing
 calculation on a detailed study of temperature and salt fraction
 dependence of conductivity of polymer-salt complexes. In this study 
 the  variation of
 $\tau_{r}$ with temperature or other factors may be important. However, as
 we have shown the rearrangement effect itself cannot be responsible for a
 change in D by as much as several orders of magnitude and the effect is
 most pronounced at low crystallinities. \newline  {\bf Acknowledgement}  
 \newline AJB is grateful to the UGC for the award of Senior Research 
 Fellowship. The authors 
 are also grateful to  K.P.N. Murthy, S.M. Bhattacharjee, T.K. Ballabh
 and T.R. Middya for stimulating discussions.

\newpage 
 {\bf References } \begin{enumerate} \item J.P. Bouchaud and A. Georges, 
Phys. Rep.
 {\bf 195} (1990) 127. \item S. Havlin and D. Ben-Avraham, Adv. in Phys 
{\bf 36}
(1987) 695. \item J.W. Haus and K.W. Kehr, Phys. Rep. {\bf 150} (1987) 263.
 \item A. Nitzan and M.A. Ratner, J. Phys. Chem. {\bf 98} (1994) 
1765.\item S.D. Druger,
 A. Nitzan and M.A. Ratner, J. Chem. Phys. {\bf 79} (1983) 3133. 
\item S.D. Druger, M.A. Ratner and A. Nitzan, Phys. Rev. B {\bf 31} (1985)
 3939. \item A.J. Bhattacharyya,
  T.R. Middya and S. Tarafdar, Pramana: J. of Phys. \newline {\bf 50(3)} 
(1998) 1.
 \item W. Voigt, Ann. Phys. {\bf 38} (1889) 573 ; A. Reuss, Z. Agnew. 
Math. Mech. {\bf 9} (1929) 49. 
 \item S.H. Chung,
K.R. Jeffrey and J.R. Stevens, J. Chem. Phys. {\bf 93} (1991) 1803. \item 
 C. Wang, Q. Liu, Q. Cao, Q. Meng  and L. Yang, Solid State Ionics {\bf 
53-56}
(1992) 1106. \item M.A. Ratner and D.F. Shriver, Chem. Rev. {\bf 88} (1988)
109 and references therein.
 \item W. Chang and G. Xu, J. Chem. Phys. {\bf 99} (1993) 2001.
  
\end{enumerate}
\newpage 
{\bf Figure Captions : } \begin{enumerate}
 \item Fig.1 : Plot of diffusion coefficient (D) versus renewal time 
$\tau_{r}$  
 at  X=0.14 for R=9.9 ($\ast$), R=8 ($\diamond$), R=6 ($\Box$) and R=1 
(o). The Calculated values using eqn.(7) are shown by continuous lines. 
 \item Fig.2  : Plot of diffusion coefficient (D) versus renewal time 
 $\tau_{r}$
at  X=0.80 for R=9.9 ($\ast$), R=8 ($\diamond$), R=6 ($\Box$) and R=1 
(o).    Calculated values using eqn. (7) are shown by continuous lines.
\item Fig.3 : Plot of diffusion coefficient (D) versus 1/$\tau_{r}$ at X=0.14 
 for R=9.9 ($\ast$), R=8 ($\diamond$), R=6 ($\Box$) and R=1 (o). The 
continuous lines are simply lines joining the points.
 \item  Fig.4 : Plot  of $\Delta$D ($\triangle$) and D$_{r}$ (o) versus X 
for a fixed R = 9.9.  \end{enumerate} 

\end{document}